\documentclass {article}
\usepackage[dvips]{graphicx}
\bigskip \bigskip
\begin{document}

\bigskip \bigskip
\centerline{\bf A VARIATIONAL PRINCIPLE FOR TWO-FLUID MODELS }\medskip \vskip 0,5 cm \centerline{\bf Sergey Gavrilyuk, Henri
Gouin  and Yurii Perepechko}\medskip
\bigskip
\centerline{S.G. and H.G.: L. M. M. T. \,    Box 322,\,  University of
Aix-Marseille} \centerline{ Avenue Escadrille
Normandie-Niemen, 13397 Marseille Cedex 20 France}\medskip
\centerline{Y.P.: United Institute of Geology, Geophysics and Mineralogy,} \centerline{Siberian Branch of the Russian Academy of Sciences, Novisibirsk 630090, Russia}\medskip
\bigskip

\centerline{\small E-mail:  henri.gouin@univ-cezanne.fr}
\bigskip

\bigskip \bigskip
\centerline{\bf Abstract}\bigskip
A variational principle for two-fluid mixtures is proposed. The Lagrangian is constructed as the difference between the kinetic energy of the mixture and a thermodynamic potential conjugated to the internal energy with respect to the relative velocity of phases. The equations of motion and a set of Rankine-Hugoniot conditions are obtained. It is proved also that the convexity of the internal energy guarantees the hyperbolicity of the one-dimensional equations of motion linearized at rest.

   \bigskip

\font\hrm=cmr8
\baselineskip25pt
\parindent20pt
\parskip18pt
\baselineskip 18pt

\medskip

 \bigskip
\noindent{\bf 1 -- Variational approach to the description
of homogenous two-velocity media}

 The variational approach to the construction of two-fluid models was used by
 many authors
(A.Bed\-ford$\ $ \& D.S. Drumheller (1978), V.L.Berdichevsky (1983),
 J.A. Geurst (1985,1986), H. Gouin
(1990)). Here we give its generalisation for
 the case of homogeneous two-fluid mixtures.  A physical example of such flow is
 a motion of a mixture
of two gases with quite  different molecular weights. The Hamilton's principle
is applied  for perfect fluid motions.  In order to obtain the equations of motion of the mixture
from this variational principle,  we neglect the dissipative effects.  We will then consider
only mechanical processes by suppressing  thermal evolution.
 We suppose  that the
homogeneous mixture motion is well represented  by the
 velocities of its components
$\displaystyle \vec u_1, \ \vec u_2$
, the average densites
$\displaystyle\rho_1, \ \rho_2$
and the total internal energy $\ U$. The total energy of a two-velocity medium
 is written in the form:
$$ E = {\rho_1\vert\vec u_1\vert^2\over 2} \ + \
{\rho_2\vert\vec u_2\vert^2\over 2} \ + \ U.
$$
In order to define the internal energy  of  one-velocity  media,
a moving coordinate system, where the elementary volume of the continuum
is at rest, is considered. The total energy of the continuum with respect to
this system is called the internal energy of the motion.
  For a two-velocity medium, there is no coordinate system, within the framework  of which any motion
could be disregarded. This is the reason why the standard definition of internal
  energy leads to its dependence on
the relative motion of components.

 Let \ $\vec w \ =  \ \vec u_2 - \vec u_1 \ , \ w = \vert \ \vec w \ \vert.$
We propose  the following extended form of Hamilton's
 principle  of least action for  the two-velocity systems:
$$ \delta IÊ= 0 \ , \
I = \int^{t_2}_{t_1} \int_{\cal D} (\rho_1 \ {\vert \vec u_1 \vert^2\over 2 } \ + \
\rho_2 \ {\vert \vec u_2 \vert^2\over 2 } \  - \
W \ (\rho_1 \ , \ \rho_2 \ , \  w )) d \vec x \ dt
\eqno(1.1)
$$
with the kinematic contraints
$$ {\partial \rho_i\over \partial t} + {\rm div} \ (\rho_i \ \vec u_i) = 0.
\eqno(1.2)
$$
Here  $\lbrack \ t_1  ,\ t_2 \rbrack $    is  a time interval ,
${\cal D}$  is a domain in the physical space, the potential
$W (\rho_1 \ , \rho_2 \ , w)$
is connected with the internal energy  $U$  by
  the partial Legendre transformation
 with respect to the variable $w$:
$$U(\rho_1,\rho_2 ,i) = W (\rho_1, \rho_2 , w)
  - w {{\partial W}\over {\partial w}} = W + i \  w , \ \ \
\hbox{\rm with}\  \ \ \ \ \
i = - {{\partial W}\over {\partial w }}.
\eqno(1.3)
$$
 The internal energy of the medium
$\displaystyle {U}  \ (\rho_1 \ , \ \rho_2  \ , \ i)$
is generally {\it a   convex}  function  of its variables.
 This  assumption corresponds to the condition of thermodynamic stability.
As a consequence,  \ $W (\rho_1 \ , \rho_2 \ , w)$ \ is a convex
function with respect to variables \ $\rho_1 \ , \rho_2$ \ and a concave
 function with respect to variable \ $w$.
A simple case is associated with
$$ W(\rho_1 \ , \ \rho_2  \ , \ w) = \varepsilon(\rho_1\ , \ \rho_2)
 \ - \ {a(\rho_1\ , \rho_2) w^2\over 2} \ ,$$ $$  \  U
 \ (\rho_1 \ , \ \rho_2  \ , \ i) = \varepsilon \ (\rho_1\ , \ \rho_2) \ +
 {i^2\over {2 \ a (\rho_1\  
\rho_2)}},
\eqno(1.4)$$
 where  $\displaystyle\varepsilon (\rho_1\ , \ \rho_2) \ $ is
a convex function with respect to \
$\rho_1\ , \ \rho_2$  \ and  \ $ a \ ( \rho_1\ , \ \rho_2 )$ \ is a positive function.
If the relative velocity $\vec w$ is small enough, the energy $U$ is convex.
The example (1.4) is a reasonable approximation  of a general
case as $W$ is an analytic function  of the velocities $\vec u_i$
(and hence, an analytic function of $w^2$ only).  The term in equation (1.4) which is
quadratic with respect to the relative velocity $\vec w$,
can be  considered as the energy due to an added mass effect. We note that this quadratic dependence
is usually used in the theory of bubbly liquids (V.L. Berdichevsky, 1983, J.A. Geurst, 1985, 1986).
To derive the governing equations and the Rankine-Hugoniot conditions, it is not necessary to focus on
the particular case (1.4).

\noindent
{\bf 2 -- Governing equations}

We introduce Lagrange coordinates \ $\vec X_i$  \ for each component:
$$ {{d_i \vec X_i}\over {dt}} = 0,\quadÊ{\rm where}Ê\quad
  {d_i\over dt} \ =  \ {\partial\over \partial t}
 +\ (\vec u_i\nabla), \ \ i = 1,2.
\eqno(2.1)$$
It follows from  (2.1)  that
$$\vec u_i \ = \ - \ ({\partial \vec X_i\over \partial \vec x})^{-1} \ <\
  {\partial \vec X_i\over \partial t} \ >,
\eqno(2.2)$$
where the operation  $A < \vec f >$  denotes the product of the tensor  $A$
 by the vector  $\vec f$.\newline
 Introducing the Lagrange multipliers
  $\displaystyle \varphi_1(t , \vec x)\ ,\ \varphi_2(t , \vec x)$ ,
corresponding to the balance of masses
  $\ (1.2)$ , we consider the Lagrangian  $L$  of the system
$$ L \ = \rho_1 \ ( \ {1\over 2 }
 \ \vert \vec u_1 \vert^2 \ - \ {d_1\varphi_1\over d t} )+ \rho_2 (\ {1\over 2 }
 \ \vert \vec u_2 \vert^2\ -  \ {d_2\varphi_2\over d t})- W (\rho_1, \rho_2 , w).
\eqno(2.3)
$$
The formulae (2.2),(2.3)  give the Lagrangian $L$
as a function of  variables
$\displaystyle {\partial\vec X_i\over \partial t} \ ,$  $\displaystyle
 \ {\partial\vec X_i\over \partial \vec x},
{\partial \ \varphi_i\over \partial t}$,
$\displaystyle {\partial
\ \varphi_i\over\partial \vec x}$,
$ \rho_1 , \rho_2.$
Calculating the corresponding variational derivatives,
 we find the governing equations of motion :
$$
{d_i \vec K_i\over d t } \ +
 \  {\partial \ \vec u_i\over \partial \ \vec x}^{*}<\vec K_i \ - \ \vec u_i> \ +
\ \nabla^{*}({\partial W\over \partial \rho_i}) \ = \ 0 \ , \eqno(2.4)$$
$$\vec K_i=\vec u_i-(-1)^i{1\over\rho_i}{\partial W\over \partial w}
{\vec w\over w}
\ ,\ i=1,2 
$$
and, of course, the mass conservation laws (1.2).
Here  and later "${*}$" denotes the transposition.
For the case (1.4)  equations (2.4) have the
form :
$${{d_i\vec K_i}\over {dt}}+(-1)^i \ {a\ (\rho_1,\rho_2)\over \rho_i}\
 {\partial\vec u_i\over \partial \vec x}^{*}<\vec w>\ +\nabla^{*}
\bigg({\partial\varepsilon\over \partial\rho_i}-{1\over 2}
 {w}^2{\partial a\over \partial\rho_i}\bigg)=0,
$$
$$
 \vec K_i=\vec u_i+(-1)^i\ {a(\rho_1,\rho_2)\over \rho_i}\ \vec w.
$$
The system (1.2) , (2.4)\  yields the momentum conservation law and the
 energy conservation law, corresponding to the homogeneity
of the Lagrangian with respect to space and time variables:
 $$ {\partial\over {\partial t}}
  \bigg( \rho_1 \vec u_1 +\rho_2\vec u_2\bigg)+
{\rm div}\ (\rho_1\vec u_1 \otimes \vec u_1^*+\rho_2\vec u_2\otimes \vec u_2^* -
{\partial W\over {\partial w}}\ {{\vec w\otimes \vec w^*}\over {w}} +
$$
$$
+( \rho_1\ {\partial W\over \partial \rho_1}+
\rho_2{\partial W\over\partial \rho_2}-W) I\bigg) =0,
\eqno(2.5)
$$
$$ {\partial\over \partial t} \ \bigg( {1\over 2}\ \rho_1 \vert
 \vec u_1
\vert ^2+ {1\over 2}\ \rho_2 \vert
 \vec u_2
\vert ^2  + U\bigg) + {\rm div}\ \bigg( \rho_1 \vec u_1 \bigg( {\vert \vec u_1
\vert ^2\over 2}+ {\partial W\over \partial \rho_1}\bigg)
+\rho_2\vec u_2\ \bigg({\vert \vec u_2 \vert ^2\over 2}+
 {\partial W\over \partial \rho_2}\bigg)-  
$$
$$
 - {\partial W\over \partial w}( \vec u_2\otimes \vec u_2^* - \vec u_1\otimes
\vec u_1^* )<{\vec w\over w}>\bigg) =0,
\eqno(2.6)
$$
where $\ \otimes\  $ denotes the tensor product and $\ I\ $ is the unit tensor.

The question of the {\it hyperbolicity} of the system (1.2),(2.4) is of great interest.
 Unfortunately,
the multi-dimensional case is not very simple and this
 is why we restrict our attention on the
one-dimensional case. Moreover, to simplify the calculations,
 we consider the case (1.4)
 and
linearize our system in the  neighbourhood of the equilibrium state  $\vec u_1^0=
\vec u_2^0 = 0\ ,\ \rho_1^0, \rho_2^0.$ Straightforward  calculations
give the following result: if $ \ \varepsilon\
(\rho_1\ ,\ \rho_2)$
is a convex function and $\ a\ (\rho_1\ ,\ \rho_2)\ >0$
then linearized at rest, the system (1.2), (2.4)
is hyperbolic.

\noindent
{\bf 3 -- The Rankine-Hugoniot conditions}

The non-linearity and hyperbolicity of the system (1.2),(2.4)
  imply  the necessity to obtain  Rankine-Hugoniot conditions across shocks.
In multi-dimensional case the conservation laws (1.2),(2.5),(2.6)
 are not sufficient to obtain the whole set of such relations.
 In what follows, we show that the
 additional jump conditions can be derived from the variational
principle.

Let us define  variations of particules deduced from the relation
 $\displaystyle \ \vec x = \vec\Phi_i\ (\vec X_i\ ,\ t\ ,\ \varepsilon _i)\ $ and
its inverse  $\displaystyle \ \vec X_i = \vec\Psi_i\ (\vec x\ ,\ t\ ,\
\varepsilon _i).$  Here  $\ \displaystyle \varepsilon _i\ ,\ i=1,2\ $
are small parameters defined in a
 neighbourhood  of zero. One defines virtual displacements
 $\displaystyle\ \delta_i\vec x\ $ and $ \ \displaystyle\ \delta_i\vec X_i\ $\ by
$$\displaystyle \delta_i \vec x\ = {\partial\vec\Phi_i\over
 \partial\varepsilon_i}(\vec X_i,t,0)\ ,\  \delta_i\vec X_i\
 = {\partial\vec\Psi_i\over  \partial\varepsilon_i}(\vec x,t,0).
\eqno(3.1)
$$
It is clear that
 $\displaystyle \delta_i\vec X_i\ = - F_i^{-1}<\delta_i\vec x>$,
where
 $\displaystyle {F_i={\partial\vec x\over  \partial\vec X_i}}$.
For any variable $\ \alpha_i\ (\vec x\ ,\ t)\ $ we define their eulerian perturbations
 $\hat \alpha_i\ (\vec x\ ,\ t\ ,\ \varepsilon_i)$ and lagrangian perturbations
$\tilde \alpha_i\ (\vec X_i\ ,\ t\ ,\ \varepsilon_i)$
$\ =\alpha_i
(\vec\Phi_i (\vec X_i, t,\varepsilon_i), t ,\varepsilon_i).$
The variations $\ \hat\delta_i \alpha_i$ \ and
$\ \tilde\delta_i \alpha_i$  \ of \ $\alpha_i$\ are defined
 by
$$\displaystyle \ \hat\delta_i \alpha_i =
{\partial \hat\alpha_i\over \partial\varepsilon_i}(\vec x,t,0),
 \ \  \ \tilde\delta_i \alpha_i = {\partial \tilde\alpha_i\over
 \partial\varepsilon_i}(\vec X_i, t,0).
$$
It follows from the above definitions  that
$$\displaystyle \hat \delta_i\alpha_i= \tilde\delta_i\alpha_i -
{\partial\hat \alpha_i\over \partial\vec x}\ < \delta_i \ \vec x >.
\eqno(3.2)
$$
 \medskip
In particular,
$$\hat \delta_i\rho_i=- {\rm div}(\rho_i \ \delta_i\vec x)\ ,\
 \hat\delta_i\vec u_i = {d_i\over dt}\ \delta_i\vec x -
{\partial \vec u_i\over  \partial \vec  x}\ < \delta_i\vec x >.
\eqno(3.3)
$$
We define
$$\displaystyle \delta_i\ I = \lim_{\varepsilon_i\rightarrow 0}\ \ {I (\varepsilon_i)
 - I (0)\over \varepsilon_i}.$$
Then
$$
\delta_i\ I=\int^{t_2}_{t_1}\ \int_{\cal D} \
(\hat\delta_i \rho_i({\vert \vec u_i\vert ^2 \over 2 }-
{\partial\ W\over \partial\rho_i})+\rho_i\vec K_i^{*}\hat\delta_i\
 \vec u_i \ )\ d\vec x\ dt.
\eqno(3.4)$$
It follows from (3.3)\ (3.4) that
$$
\displaystyle \delta_i\ I = \int_{t_1}^{t_2}\  \ \int_{\cal D}\bigg(-{\rm div}
(\rho_i\delta_i\vec x)({\vert \vec u_i\vert ^2\over 2}-{\partial W\over
 \partial \rho_i})+\rho_i\ \vec K_i^{*}({d_i\over dt}\ \delta_i\vec x-
{\partial \vec u_i\over \partial \vec  x}<\delta_i\ \vec x>)\bigg)\ d\vec x \ dt  
$$
$$\displaystyle = \int_{t_1}^{t_2}\ \ \int_{\cal D}\ \bigg(\ -\rho_i \delta_i\vec x^*\bigg(\vec K_{it}+{\partial \vec K_i\over \partial\vec x}<\vec u_i>+{\partial\vec u_i^*\over \partial\vec x}<\vec K_i> + \nabla^*({\partial W\over \partial\rho_i}-{\vert \vec u_i\vert ^2\over 2})\bigg)+$$
$$\displaystyle +{\partial\over \partial\ t}(\rho_i\delta_i\vec x^*\
\vec K_i)+{\rm div}(\rho_i\vec u_i(\delta_i\vec x^*\ \vec K_i)+\rho_i\delta_i
\vec x({\partial W\over \partial\rho_i}-
{\vert \vec u_i\vert ^2\over  2}))\bigg)\ d\vec x\ dt.$$
As a consequence, we obtain equations of motion (2.4)
 and the jump conditions:
$$ \bigg[\ \rho_i (\vec n^*\ \delta_i\vec x)\
({\partial W\over \partial\rho_i}-{\vert \vec u_i\vert ^2\over 2})+\rho_i\
(\vec n^*\ \vec u_i)\ (\vec K_i^*\ \delta_i\vec x) - D_n\ \rho_i(\vec K_i^*
\delta_i \vec x)\bigg]\ =0\ ,
\eqno(3.5)
$$
where $\displaystyle \vec n\ $  is the unit normal vector to the shock surface
and $\ D_n\ $    is the  normal velocity of the shock.
Since $\ \displaystyle \delta_i\vec x\ $  is not continuous
 across the shock, it is not straightforward  to obtain Rankine-
Hugoniot  conditions in terms of desired quantities. Nevertheless, taking into
account that across the shock
$$
[\rho_i F_i^*<\vec n>] = 0,
$$
such difficulties can be overcame. We finally obtain from (3.5):
$$
 \bigg[\ {\partial W\over  \partial\rho_i}-
{\vert \vec u_i\vert ^2\over 2}+\vec K_i^*\ \vec u_i -
 D_n\ (\vec K_i^* \vec n)\bigg]\ =0\ ,
 \eqno(3.6)
$$
$$
[\ \vec K_i-(\vec K_i^* \ \vec n)\vec n\ ]\ =0.
\eqno(3.7)
$$
In the limiting case when  the velocities of the components coincide,
  these conditions
 reduce to
 the conservation of the tangential component of
the velocity and  the conservation of the Bernoulli constant
across the shock.
We note also that
conditions (3.6), (3.7) are obtained from the variational principle
 (1.1) without any assumption
on the flow properties. But it can be shown directly that they correspond to the jump
 conditions for the additional conservation laws admitted by the system  (1.2), (2.4) :
$$ {\rm rot}\vec K_i=0,\quad
 {\partial\vec K_{i}\over \partial t} +\nabla^*({\partial W\over \partial\rho_i}-
{1\over 2}\vert u_i\vert ^2 + \vec K_i^*\ \vec u_i)=0.
\eqno(3.8)
$$
The conservation laws (2.5), (2.6) also imply momentum balance and energy balance at the
shock.  Finally note that
all possible jump conditions are of great interest but the correct choice
 of the jump conditions
depends on the physics of  the problem.
\vskip 0.25cm
\bigskip
\noindent{\bf References}

\medskip
  Berdichevsky V.L. (1983) Variational principles of continuum mechanics. Moscow : Nauka, 1983.
  
 Bedford A. \&  Drumheller D.S. (1978)
A variational theory of immiscible mixtures.  {\it Arch. Rat. Mech. Annal.} 1978. Vol. 68. P.37-51.

  Geurst J. A. (1985) Virtual mass in two-phase bubbly flow. {\it Physica. A. }1985. Vol.129A. P.233-261.

  Geurst J. A. (1986) Variational principles and two-fluid hydrodynamics of
 bubbly liquid / gas mixtures. {\it Physica. A. }1986. Vol.135A. P.455-486.

  Gouin H. (1990) Variational theory of mixtures in continuum mechanics. {\it
 Eur. J. Mech, B / Fluids}. 1990. Vol.9, No.5. P.469-491.
\vskip 1cm

\end{document}